# Stripe-Based Fragility Analysis of Concrete Bridge Classes Using Machine Learning Techniques

Sujith Mangalathu, Ph.D.[1] and Jong-Su Jeon, Ph.D.[2]*

[1]Post-Doctoral Fellow, Department of Civil and Environmental Engineering, University of California, Los Angeles, CA 90095, USA. Email: sujithmangalath@ucla.edu

[2]Assistant Professor, Department of Civil Engineering, Andong National University, Andong, Gyeongsangbuk-do 36729, Republic of Korea. Email: jsjeon@anu.ac.kr

* Corresponding author

**Abstract:**

A framework for the generation of bridge-specific fragility curves utilizing the capabilities of machine learning and stripe-based approach is presented in this paper. The proposed methodology using random forests helps to generate or update fragility curves for a new set of input parameters with less computational effort and expensive re-simulation. The methodology does not place any assumptions on the demand model of various components and helps to identify the relative importance of each uncertain variable in their seismic demand model. The methodology is demonstrated through the case study of a multi-span concrete bridge class in California. Geometric, material and structural uncertainties are accounted for in the generation of bridge models and fragility curves. It is also noted that the traditional lognormality assumption on the demand model leads to unrealistic fragility estimates. Fragility results obtained the proposed methodology curves can be deployed in risk assessment platform such as HAZUS for regional loss estimation.

**INTRODUCTION**

Past earthquakes have demonstrated that highway bridges, as key components of transportation networks, are one of the most vulnerable components. The state of the bridge after an earthquake is critical in deciding the emergency or ordinary traffic. The likelihood of bridge damage during a seismic event can be obtained through fragility curves. Fragility curves are conditional probability statements that give the likelihood that a structure or component of that structure will meet or exceed a certain level of damage for a given ground motion intensity measure (IM).

Extensive studies have been carried out to determine the fragility analysis of highway bridges using analytical methodology (e.g., Nielson 2005, Mackie and Stojadinović 2006, Padgett 2007, Banerjee and Shinozuka 2008, Zhang and Huo 2009, Vosooghi and Saiidi 2010, Ramanathan 2012, Ghosh 2013, Mangalathu et al. 2015, Monterio 2016, Mangalathu et al. 2016, Mangalathu 2017, Monterio et al. 2017). Existing analytical fragility methodology consists of convolving demand models with capacity models. The prevalent analytical approaches in the generation of bridge fragilities are cloud approach, stripe, and incremental dynamic analysis approach (IDA). All three approaches typically employ nonlinear time history analysis of bridge models to extract seismic responses (demands). In the cloud approach, the probabilistic demand models are established by linear regression of engineering demand parameters (EDPs) and IM in a lognormal space. In the stripe analysis, ground motions are scaled to the same intensity level to find the probability distribution of EDPs. In the IDA approach, ground motions are successively scaled until the significant reduction of primary load-bearing elements (collapse prevention) in the structural system (Vamvatsikos and Cornell 2002). A comparison of the three approaches in the estimation of seismic demand model is given in Mackie and Stojadinović (2005).

All the three methods are used to determine the median (or mean) relationship between EDP and IM, and the associated uncertainty. As noted by recent researches (Jeon et al. 2017a, Mangalathu et al. 2018), single-parameter fragility curves have many drawbacks; single-parameter fragility curves (1) demand expensive re-simulations, if there is an update in the input parameter from a field observation or new database, (2) cannot identify the relative importance of each uncertain input variable on the fragility curves, and (3) need the grouping of bridges that have statistically similar performance prior to performing fragility analysis. Also, the predictive capability of the single-parameter demand model is questioned by Mangalathu et al. (2018), especially at high IMs. At high IMs, bridge components would experience nonlinear behavior, resulting in statistical errors in their demand models because IM is the only conditioned variable (ignorance of material nonlinearities). Also, all the aforementioned methods place a lognormal assumption on the demand model. However, Karamlou and Bochini (2015) noted that this assumption introduces a significant error in the fragility curves and associated loss estimation.

This paper presents a methodology to generate seismic fragility curves for regional risk assessment by exploiting the advantages of stripe approach and machine learning techniques. Compared to existing stripe approach, this paper suggests a multi-parameter fragility methodology using Random Forests (RF) and the proposed methodology does not place any assumption on the demand model and dispersion. RF is a popular machine learning technique and is widely used in the field of statistics (Breiman 2001, Friedman et al. 2001), and bio-informatics (Díaz-Uriarte and De Andres 2006, Svetnik et al. 2003). RF has substantial advantage over other machine learning techniques because of their flexibility, intuitive simplicity and computational efficiency. Compared to other machine learning techniques (such as Lasso, Ridge, Naïve Bayes, simple neural networks), RF is not placing any strong assumption on the mapping function and is non-parametric (Friedman et al. 2001). The non-parametric approach is of special significance as recent studies noted the inadequacy of log-normal assumptions in the generation of fragility curves (Karamlou and Bochini 2015). Although RF has a superior performance in comparison to other machine learning techniques such as support vector machines and neural networks (Breiman 2001), it is not fully explored in the field of structural engineering. Rokneddin (2013) explored the application of RF for network reliability analysis of highway bridges. The author pointed out that using RF as a surrogate model can reduce expensive simulations (computational efforts). Tesfamariam and Liu (2010) investigated the application of RF and other machine learning techniques in classifying the damage state of buildings from past earthquakes. However, the capability of RF in the generation of multi-parameter fragility curves using the stripe approach is not yet explored and this paper is directed towards that.

The proposed bridge methodology is used to suggest fragility curves for three-span and four-bridges with seat abutment bridge classes in California. Although these bridges occupy more than 40% of California bridge inventory, very few studies have explored the vulnerability of these bridges. As noted in the work of Mangalathu et al. (2017) based on a statistical analysis, three-span and four-span bridges have statistically similar performance. Geometric, structural, and material uncertainties are accounted for in the current study to generate fragility curves of the selected bridge classes. A brief description of RF is given in the following section. Then, the description of the numerical modeling of the selected bridges and the associated uncertainties are provided. After that, this paper outlines the proposed methodology using RF, compare the fragility results from RF and existing stripe methodology, check normality assumptions, and identify the relative importance of uncertain input variables.

**Random Forest**

Random forest (RF) is a learning method consists of an ensemble of tree-structures. RF takes advantage of two powerful machine learning techniques: bagging and random feature selection (Breiman 2001). In bagging, each tree is independently constructed using a bootstrap sample of the training data, and the mean value of the outputs of the trees is used for prediction (Breiman 1996). RF is a revised version of bagging. Instead of using all features, RF randomly selects a subset of features to be split at each node when growing a tree. The addition of randomness makes RF perform well compared to other machine learning techniques such as support vector machines and neural networks and is robust against overfitting (Breiman 2001, Liaw and Wiener 2002). Interested readers are directed to the references (Breiman 2001, Friedman et al. 2001) for a more detailed description of RF and a general algorithm is as follows:

(1) Generate $n_t$ bootstrap samples from the training dataset.

(2) Generate a decision tree from each bootstrap sample by selecting the best split among the dataset.

(3) Predict the output of a new dataset by averaging the aggregate of predictions of $n_t$ decision trees.

The output of the RF prediction can be expressed as:

$$\hat{f}_{RF}^{n_t}(x) = \frac{1}{n_t} \sum_{1}^{n_t} f_{n_t}(x) \qquad (1)$$

where $\hat{f}_{RF}^{n_t}(x)$ denotes the outcome of RF prediction (average value) from a total of $n_t$ trees, and $f_{n_t}(x)$ is the individual prediction of a tree for an input vector *x*. The variance of the average of $n_t$ random variables with a correlation coefficient $\rho$ and standard deviation $\sigma$ is (Friedman et al. 2001):

$$\text{var}_{n_t} = \rho\sigma^2 + \frac{1-\sigma}{n_t}\sigma^2 \qquad (2)$$

An estimate of the error rate can be obtained from RF by the following steps:

(1) Predict the data which is there in the original dataset and not in the bootstrap sample (*out-of-bag*, or OOB, data) using the tree grown with the bootstrap sample for each bootstrap iteration.

(2) Aggregate the OOB predictions. Calculate the error rate, and call it the OOB estimate of error rate.

A schematic representation of RF for two variables $x_1$ and $x_2$ to assign the predictions A, B, C, D or E is given in Fig. 1. The tree initially divides the region based on the variable $x_1$, if $x_1 \geq a_1$, the left branch of the tree is activated and it assigns the predictor A or B depending on the variable $x_2$, i.e., whether $x_2 \geq a_2$. For the cases where $x_1 > a_1$, the assignment of predictor C, D or E depends on the value $a_3$ for $x_2$ and $a_4$ for $x_1$. RF creates lots of decision trees and the average prediction from each decision tree yields the output.

**Numerical Modeling of Selected Bridges, Uncertainties, and Ground Motions**

The current study selected three-span and four-span bridges in California constructed after 1970 to demonstrate the proposed fragility methodology. Ramanathan (2012) indicated that these bridge types occupy more than 40% of the California box-girder bridge inventory. Three-dimensional finite element models of these bridge types are developed in OpenSees (Mazzoni et al. 2006) with realistic representations for abutments, deck, columns, foundations, bearing, and pounding. Deck is modeled using a linear element and is connected to columns with rigid links. Displacement-based beam-column elements with fiber-defined cross-sections comprising fibers of confined and unconfined concrete and longitudinal reinforcement are used to model the columns (Fig. 2). Each column is equally divided into nine beam-column elements along the clear height of the columns, each of which has five integration points. Elasto-plastic model is used to model the bearings and the pounding effect is simulated using an inelastic compression element with the gap (Muthukumar and DesRoches 2006).

Earth pressure comprises of two types of resistance: active resistance when the abutment wall moves away from the backfill and the passive resistance when the abutment wall compress the backfill. The active resistance is assumed to be provided only by the piles (Ramanathan 2012) while the passive resistance is contributed by the soil and the piles. The hyperbolic soil model proposed by Shamsabadi et al. (2010) is used to simulate the passive response of the abutment. The response of the abutment piles is simulated using a trilinear material model presented in Mangalathu et al. (2016). Exterior shear keys are simulated following the backbone curves of experimental results by Silva et al. (2009). The foundations are modeled using linear translational and rotational springs. The various components previously described are combined to generate the bridge system model for seismic fragility analysis, and the spring connection for the various components is also shown in Fig. 2. Interested readers are directed to Jeon et al. (2017b) for a more detailed description of the modeling and numerical models for various bridge components.

To include possible uncertainties in the creation of bridge models, different sources of uncertainties, such as geometry, material, and system, are accounted for in the current study. Note that the input variables are determined in the current study on the basis of the numerical modeling technique for various bridge components and the insights from the sensitivity study on bridge demand models (Mangalathu et al. 2018). The distribution and statistical parameters of the input variables are identified based on an extensive plan review of over 1,000 randomly selected bridges in California. A more detailed description of the plan review and input parameters are provided in Mangalathu (2017). Table 1 presents the mean value (μ), standard deviation (σ), and the associated probability distribution of various input variables used in the current study.

The current study selects the suite of ground motions developed by Baker et al. (2011) which was developed as part of the PEER Transportation Research Program for the seismic risk assessment of infrastructure systems in California. Ramanathan (2012) conducted an extensive study on the suitable IM for the bridge in California. Based on the author's recommendation, spectral acceleration at 1 sec ($S_{a-1.0}$) is adopted as the IM in the current study.

Statistically significant yet nominally identical three-dimensional bridge models are created by sampling across the range of parameters presented in Table 1 using Latin Hypercube Sampling (LHS). Compared to pure random sampling using naïve Monte Carlo simulation, LHS provides an effective scheme to cover the probability space of the random variables (McKay et al. 1979). The generated bridge models are randomly paired with the selected suite of ground motions to obtain the bridge-ground motion pair for NLTHA. The two orthogonal components of the ground motions are randomly assigned to the longitudinal and transverse direction of the bridge axis. The various EDPs and the associated limit states of bridge components are presented in Table 2. *βc* is done in a subjective manner due to lack of sufficient information, and is adopted as constant across the components and the respective damage states. Also, the adopted *βc* value is a good estimate for columns based on the column data base summarized by Mangalathu (2017).

**Proposed Fragility Framework**

*Existing Stripe-Based Fragility Methodology*

In the existing stripe-based fragility methodology, ground motions are scaled to the same IMs and perform NLTHA on the bridge models using the scaled ground motions. Lognormal distribution is fitted on the EDPs obtained from NLTHA at a single IM and is convolved with capacity models to calculate the probability of failure at the IM. The fragility function computing a failure probability can be written as:

$$P[D \geq C \mid IM = x] = 1 - F(D_{LS} \mid IM = x) \quad (3)$$

where $F(D_{LS}|IM=x)$ is the cumulative probability of obtaining a specified limit state ($D_{LS}$) at $IM = x$. If both the demand and capacity models follow lognormal distributions, this cumulative distribution function for a specified IM can be expressed as (see Fig. 3):

$$F(D_{LS} \mid IM = x) = \Phi\left[\frac{\ln D_{LS} - \lambda}{\beta}\right] = \int_0^{D_{LS}} \frac{1}{\sqrt{2\pi}\beta z} \exp\left[-\frac{1}{2}\left(\frac{\ln z - \lambda}{\beta}\right)^2\right] dz \quad (4)$$

where $\Phi[\bullet]$ is the cumulative normal distribution function, and

$$\lambda = \ln\left(\frac{\mu}{\sqrt{1 + v_x^2}}\right) \quad (5)$$

where $\mu$ is the mean of response data and $v_x$ is the coefficient of variation. The overall uncertainty $\beta$ (dispersion) can be defined as the square root of the sum of the squares of response uncertainty due to record-to-record variation ($\beta_{D|IM}$) from the demand model and limit state uncertainty ($\beta_C$):

$$\beta = \sqrt{\beta_{D|IM}^2 + \beta_C^2} \quad (6)$$

where

$$\beta_{D|IM} = \sqrt{\ln(1 + v_x^2)} \quad (7)$$

The failure probability for a limit state for each IM obtained from Eq. (3) through Eq. (5) is used to construct a fragility curve for the limit state over the entire range of IMs, as illustrated in Fig. 3. This fragility method has an ability to develop fragility curves with different dispersions at each IM level. However, this method should satisfy the normality assumption at each level of IMs with an acceptable margin of error.

*Proposed Stripe-Based Fragility Methodology Using Random Forest*

A new fragility methodology utilizing the features of RF and stripe-method is suggested in this section. The proposed stripe-based fragility methodology has several advantages compared to the traditional stripe-based fragility method such as:

(1) The proposed methodology can identify the relative importance of variables at each IM. The identification of the relative importance of uncertain parameters helps bridge owners to spend their resources judiciously in updating the input parameter database for future fragility analysis.

(2) The methodology can be used to update the fragility curves without expense re-simulation if the input parameters are required to be updated in future.

(3) The proposed methodology does not place any assumptions on the demand model as RF-based demand model is non-parametric.

(4) Although the proposed methodology consists of sampling the bridge models across the range of uncertain input parameters using LHS, the trained RF model can be later used to develop fragility curves for a fixed set of input parameters (called bridge-specific fragility curves).

The steps involved in the proposed methodology are outlined below:

**Step 1:** Conduct NLTHAs of bridge models using Latin Hypercube sampling technique and scale ground motions to the desired IMs.

**Step 2:** Establish a predictive model connecting the input parameters (*IM*, $T_m$, and modeling parameters in Table 1) and output parameter (demand) using RF at each IM. Step 2 helps identify the relative importance of the uncertain input variables at each IM. Note that the predictive model is established by taking logarithms for the input and output data to reduce the nonlinearity in the relationship between the input and output parameters (Jeon et al. 2017a).

**Step 3:** Generate a large number of demand estimates (*N*, 1 million in the current study) for each component, $k_i$, using their respective RF demand model by generating *N* values of input parameters randomly generated based on their probabilistic distribution. If the fragility curve is intended for a specific bridge, establish the demand model by only accounting for the material nonlinearity and $T_m$.

**Step 4**: Generate *N* capacity values for a specific damage state for each bridge component based on the assumed distribution of the limit states (Table 2).

**Step 5**: Obtain the probability of failure ($p_{f,IM=x}$) by comparing the capacity values (Step 4) with the demand values (Step 2). i.e., $p_{f,IM=x} = \dfrac{N_f}{N}$ where $N_f$ is the number of samples where the demand value is greater than the capacity value.

**Step 6:** Repeat Steps 2 and 5 for different *IMs* and for different bridge components to construct fragility curves, as illustrated in Fig. 3.

Although there is no assumption placed on the demand model, the capacity model is assumed as lognormal in the current study following the previous work on bridge fragilities (Padgett 2007; Ramanathan 2012; Mangalathu 2017). Developing bridge-attribute related capacity model needs extensive experimental data and is beyond the scope of the current study. Also, Mangalathu (2017) suggested that lognormal-fit is the best possible fit for capacity models based on available experimental results.

**Fragility Estimates of Selected Bridge Classes**

*Performance Evaluation of Various Machine Learning Methods*

The predictive capability of parametric machine learning techniques such as lasso regression and elastic net (EN) and non-parametric machine learning techniques such as support vector machine (SVM) and RF is initially compared to evaluate the performance of RF in comparison to other machine learning techniques. Table 3 presents the coefficient of determination ($R^2$) and mean square error (MSE) for various regression techniques at various IM levels in the case of the column curvature ductility (COL) and bearing displacement (BRG). RF has the best predictive capability with high $R^2$ and low MSE in comparison to other methods across all IM levels (marked as bold in the table). Although the results are shown here only for two components, the observation is valid for other components as well. The following section examines the validity of the log-normality assumption placed on the demand model.

*Checking Normality Assumption in Traditional Stripe Analysis*

Fig. 4 shows the histogram and probability plot of $\mu_\phi$ and $\delta_b$ at $S_{a-1.0}$ = 1.0g. To investigate the validation of traditional lognormality assumption (or the normality assumption on the log-transformed variables) placed on the demand model in the strip analysis, Kolmogorov–Smirnov test (Vidakovic 2011) is carried out, which checks the null hypothesis that the data are lognormally distributed. If the *p*-value is less than the cut-off value of 0.05, the null hypothesis is rejected. Thus, there is enough evidence that the data do not follow a lognormally distributed population. The *p*-values for $\mu_\phi$ and $\delta_b$ are close to zero, and thus the data do not follow a lognormal distribution. Although not shown here, the lognormality assumption is not true for all the EDPs considered in the current study. Therefore, use of lognormality assumption leads to unrealistic demand models. As mentioned before, RF is not placing any assumption on the demand models.

### *Comparison of Traditional and Proposed Methodology*

Fig. 5 compares component fragility curves for the selected bridge classes (combined three-span and four-span bridges with the uncertainties mentioned in Table 1) using the existing stripe-method and the proposed RF-based methodology. It is seen from Fig. 5 that the dispersion (or lognormal standard deviation) associated with the proposed method is less compared to the existing strip method in that the slope for the proposed method is stiffer. The observation is valid for all the bridge components at all the limit states under consideration. It is also noted that there is not much statistical difference between the median values of the fragilities obtained from two methods at various limit states for the selected bridge components.

### *Relative Importance of Uncertain Parameters at Each IM*

Depending on the number of the input variables used in the decision trees, RF can be used to estimate the relative importance of the uncertain input variables. RF estimates the relative importance of a variable by noting the increase in the OOB error of the variable for different permutations while the other variables are kept constant. Fig. 6 shows the relative importance of the uncertain variables considered in the current study for three different IMs: $S_{a-1.0}$ = 0.2g, 0.6g, and 1.0g. The relative importance of the input variables for all the EDP's at $S_{a-1.0}$ = 0.6g is given in Fig. 6. Figs. 6 and 7 should be interpreted as the relative importance of the variables in the estimation of the seismic demand model given the uncertainties reported in Table 1. Following inferences can be obtained from Figs. 6 and 7.

(1) Span length ($L_m$), longitudinal reinforcement ratio ($\rho_l$), height of the column ($H_c$) approach span to main span ratio ($\eta$), and deck width ($D_w$) are the variables that have a significant influence on the seismic demand of bridge components for all the EDPs.

(2) Depending on the EDP under consideration, different variables have different influence on the demand model. For example, the abutment pile stiffness ($K_p$) has a significant influence on the abutment response in passive, active, and transverse directions (Fig. 7) while $K_p$ does not have a significant influence on the column curvature ductility.

(3) The relative importance of the variables changes with the change in IM. For example, Fig. 6 shows the relative importance of $L_m$ is higher at $S_{a-1.0}$ = 0.6g in comparison to $S_{a-1.0}$ = 0.2g and 1.0g.

(4) Fig. 6 underscores the need to include various uncertainties in the seismic demand model, if the fragility curves are intended for a regional risk assessment. The input parameter such as $L_m$, $\rho_l$, $H_c$, $\eta$, and $D_w$ have significant influence on the demand model and neglecting the uncertainty in these parameters leads to the unreliable fragility curves.

(5) The relative importance helps quantify the associated error if the uncertainties associated with a specific input parameter are not properly evaluated in the estimation of seismic vulnerability. For example, the error in the uncertainty estimation of $m_f$ might have a minimal impact on the fragility curves, while the error in the uncertainty estimation of $L_m$ leads to unrealistic fragility curves.

The influence of $L_m$ and $\eta$ on the demand models can be attributed to the fact that $L_m$ and $\eta$ increase the mass and flexibility of deck (bridge system), leading to the increase of seismic demands. $\mu_\phi$ is significantly affected by $D$, $H_c$, and $\rho_l$ at all the IMs. The strength and stiffness of columns are a function of $D$, $H_c$, and $\rho_l$, which explains the relative importance of $D$, $H_c$, and $\rho_l$ in $\mu_\phi$. Abutment response is significantly influenced by the back fill soil type ($BT$), abutment height ($H_a$), and abutment pile stiffness ($K_p$). This is due to the fact that force-displacement relation of abutments is a function of these variables.

*Component Vulnerability of Selected Bridge Configurations*

Fragility curves are generated for three-span and four-span bridge configurations using the proposed methodology reflecting the uncertainties listed in Table 1. Fig. 8 shows the comparison of the fragility curves for three-span and four-span bridges for the moderate damage state. It is noted that there is not much statistical difference between the fragilities of three-span and four-span bridges. This observation is consistent with the previous work on grouping the bridge classes based on the seismic demand (Mangalathu 2017, Mangalathu et al. 2017).

To implement in regional risk assessment platform such as HAZUS (for practical use), the developed RF-based fragility curves in Fig. 5 are defined as a lognormal cumulative distribution function with median (*λ*) and dispersion (*β*). The lognormal cumulative distribution function is fitted to the discrete points of failure probability by minimizing the sum of the square of residuals between the actual and fitted values. Fig. 9 shows the difference between the actual and fitted fragility curves for all the EDPs of bridge classes (combined three-span and four-span) at the moderate ($LS_2$) and complete ($LS_3$) damage states. This comparison indicates that both curves are almost identical for all the EDPs and all limit states. The median and dispersion for all components across four limit states are provided in Table 4. The fragility values reported in the table can be used for regional risk and loss estimation of three-span and four-span bridges.

**CONCLUSIONS**

This paper presents a methodology to generate bridge-specific fragility curves utilizing the capabilities of stripe approach and machine learning technique. Unlike the existing methodologies, this methodology does not place any assumption on the seismic demand and therefore provides more reliable fragility curves for regional risk assessment of bridges. The methodology is demonstrated by generating fragility curves for three-span and four-span bridges in California accounting for the material, structural, and geometric uncertainties.

Numerical models including the material and geometric uncertainties are created in OpenSees and are paired with scaled ground motions. Nonlinear time history analysis is carried to estimate the seismic demand of bridge model-ground motion pairs. A machine learning technique called Random Forest is used to establish a predictive equation connecting the uncertain input parameters and output (demand parameter). Various demand parameters such as column curvature ductility, abutment displacement in the passive, active, and transverse direction, superstructure unseating displacement, and elastomeric bearing displacement are considered in the current study. The proposed prediction model can be used to estimate the demand parameters for a new set of uncertain parameters. The demand values are compared with capacity values to generate the fragility curves. The proposed demand model can be thus used to estimate the seismic demand for a new set of input parameters without expensive simulation. It is also noted that the traditional lognormal assumption on the seismic demand model leads to unrealistic fragility curves.

The proposed methodology also helps to identify the relative importance of uncertain input parameters on the seismic demand model of various bridge components. It is noted that the span length, approach span-to-main span ratio, longitudinal reinforcement ratio, deck width, and column height are the variables that have a significant influence on the seismic demand (and fragilities) of all bridge components. Using the proposed methodology, fragility curves are generated for three-span and four-span bridges in California. The suggested fragility curves can be implemented in risk assessment platform such as HAZUS for a more accurate and reliable seismic loss estimation. Although the methodology is demonstrated for some selected bridge classes, the methodology is relevant and applicable to other structural systems.

**Acknowledgements**

This research was supported by Basic Research Program in Science and Engineering through the National Research Foundation of Korea funded by the Ministry of Education (NRF-2016R1D1A1B03933842).

**List of Tables**

**Table 1**. Uncertainty Distribution Considered in the Bridge Models (Mangalathu 2017)

| Parameter | Type[§] | Parameters | | Truncated limit | |
|---|---|---|---|---|---|
| | | Mean ($\mu$) | Standard deviation ($\sigma$) | Lower | Upper |
| **Superstructure (prestressed concrete)** | | | | | |
| Number of spans, $N_s$ (3-span versus 4-span) | B | – | – | – | – |
| Main-span length, $L_m$ (m) | N | 47.24 | 13.72 | 22.86 | 76.20 |
| Ratio of approach-span to main-span length, ($\eta = L_s/L_m$) | N | 0.75 | 0.2 | 0.4 | 1.0 |
| Deck width (5 cells), $D_w$ (m) | N | 17.37 | 2.44 | 15.24 | 20.12 |
| **Interior bent** | | | | | |
| Concrete compressive strength, $f_c$ (MPa) | N | 31.37 | 3.86 | 22.75 | 39.09 |
| Rebar yield strength, $f_y$ (MPa) | N | 475.7 | 37.9 | 399.9 | 551.6 |
| Column clear height, $H_c$ (m) | LN | 7.13 | 1.15 | 5.18 | 9.75 |
| Column longitudinal reinforcement ratio, $\rho_l$ | U | 0.02 | 0.006 | 0.01 | 0.03 |
| Column transverse reinforcement ratio, $\rho_t$ | U | 0.009 | 0.003 | 0.004 | 0.013 |
| **Deep foundation (pile group, two-column bent)** | | | | | |
| Translational stiffness, $K_{ft}$ (kN/mm) | LN | 175.1 | 0.44 | 70.05 | 437.8 |
| Transverse rotational stiffness, $K_{fr}$ (GN-m/rad) | LN | 1.36 | 0.28 | 0.54 | 3.39 |
| Transverse/longitudinal rotational stiffness ratio, $k_r$ | LN | 1.0 | 1.5 | 0.67 | 1.5 |
| **Exterior bent (seat-type abutment on piles)** | | | | | |
| Abutment backwall height, $H_a$ (m) | LN | 3.59 | 0.65 | 2.90 | 6.10 |
| Pile stiffness, $K_p$ (kN/mm) | LN | 0.124 | 0.045 | 0.058 | 0.234 |
| Backfill type, $BT$ (sand vs. clay) | B | – | – | – | – |
| **Bearing (elastomeric bearing)** | | | | | |
| Stiffness per deck width, $K_b$ (N/mm/mm) | LN | 0.630 | 0.299 | 0.230 | 1.436 |
| Coefficient of friction of bearing pad, $\mu_b$ | N | 0.3 | 0.1 | 0.1 | 0.5 |
| **Gap** | | | | | |
| Longitudinal (pounding), $\Delta_l$ (mm) | LN | 23.3 | 12.4 | 7.6 | 55.9 |
| Transverse (shear key), $\Delta_t$ (mm) | U | 19.1 | 11.0 | 0 | 38.1 |
| **Other parameters** | | | | | |
| Mass factor, $m_f$ | U | 1.05 | 0.06 | 0.95 | 1.15 |
| Damping ratio, $\xi$ | N | 0.045 | 0.0125 | 0.02 | 0.07 |
| Acceleration for shear key capacity (g), $a_s$ | LN | 1 | 0.2 | 0.8 | 1.2 |
| Earthquake direction (fault normal FN vs. parallel FP), $ED$ | B | – | – | – | – |
| Earthquake mean time period, $T_m$ (s) | LN | 0.607 | 0.47 | 0.16 | 1.78 |

[§] N = normal, LN = lognormal, U = uniform, and B = Bernoulli distribution.

**Table 2.** EDPs and Limit State Models of Various Bridge Components (Mangalathu 2017)

| Component | Median value, $S_c$ | | | | Dispersion, ($\beta_c$) |
| --- | --- | --- | --- | --- | --- |
| | Slight (LS$_1$) | Moderate (LS$_2$) | Extensive (LS$_3$) | Complete (LS$_4$) | |
| Column curvature ductility (COL) | 1 | 5 | 8 | 11 | 0.35 |
| Passive abutment response (ABP, mm) | 76 | 254 | – | – | 0.35 |
| Active abutment response (ABA, mm) | 38 | 102 | – | – | 0.35 |
| Transverse abutment response (ABT, mm) | 25 | 102 | – | – | 0.35 |
| Bearing displacement (BRG, mm) | 25 | 102 | – | – | 0.35 |
| Superstructure unseating (UST, mm) | – | – | 254 | 381 | 0.35 |

**Table 3.** Comparison of $R^2$ and MSE for Various Machine Learning Methods

| EDP | Error measure | Method | $S_{a-1.0}$ | | | | | | | | | | | |
| --- | --- | --- | --- | --- | --- | --- | --- | --- | --- | --- | --- | --- | --- | --- |
| | | | 0.1g | 0.2g | 0.3g | 0.4g | 0.5g | 0.6g | 0.7g | 0.8g | 0.9g | 1.0g | 1.1g | 1.2g |
| COL | $R^2$ | Lasso | 0.57 | 0.62 | 0.64 | 0.69 | 0.72 | 0.73 | 0.73 | 0.72 | 0.72 | 0.69 | 0.68 | 0.67 |
| | | EN | 0.56 | 0.61 | 0.63 | 0.69 | 0.72 | 0.73 | 0.73 | 0.72 | 0.72 | 0.69 | 0.68 | 0.66 |
| | | SVM | 0.53 | 0.58 | 0.60 | 0.67 | 0.70 | 0.71 | 0.71 | 0.70 | 0.70 | 0.67 | 0.65 | 0.64 |
| | | RF | **0.76** | **0.80** | **0.83** | **0.84** | **0.85** | **0.85** | **0.86** | **0.85** | **0.85** | **0.83** | **0.82** | **0.81** |
| | MSE | Lasso | 0.11 | 0.17 | 0.23 | 0.22 | 0.20 | 0.19 | 0.18 | 0.18 | 0.17 | 0.18 | 0.19 | 0.19 |
| | | EN | 0.11 | 0.17 | 0.23 | 0.22 | 0.20 | 0.19 | 0.18 | 0.18 | 0.17 | 0.18 | 0.19 | 0.19 |
| | | SVM | 0.12 | 0.19 | 0.26 | 0.23 | 0.21 | 0.20 | 0.20 | 0.19 | 0.18 | 0.19 | 0.20 | 0.21 |
| | | RF | **0.06** | **0.09** | **0.11** | **0.12** | **0.10** | **0.10** | **0.10** | **0.09** | **0.09** | **0.10** | **0.10** | **0.11** |
| BRG | $R^2$ | Lasso | 0.66 | 0.67 | 0.63 | 0.66 | 0.61 | 0.60 | 0.57 | 0.55 | 0.55 | 0.55 | 0.54 | 0.52 |
| | | EN | 0.65 | 0.66 | 0.63 | 0.65 | 0.60 | 0.60 | 0.56 | 0.54 | 0.54 | 0.54 | 0.53 | 0.51 |
| | | SVM | 0.64 | 0.65 | 0.59 | 0.63 | 0.58 | 0.58 | 0.53 | 0.51 | 0.51 | 0.52 | 0.51 | 0.47 |
| | | RF | **0.79** | **0.81** | **0.80** | **0.80** | **0.78** | **0.78** | **0.79** | **0.78** | **0.77** | **0.76** | **0.75** | **0.73** |
| | MSE | Lasso | 0.07 | 0.06 | 0.07 | 0.07 | 0.07 | 0.08 | 0.09 | 0.09 | 0.10 | 0.12 | 0.13 | 0.14 |
| | | EN | 0.07 | 0.06 | 0.07 | 0.07 | 0.07 | 0.08 | 0.09 | 0.10 | 0.10 | 0.12 | 0.13 | 0.15 |
| | | SVM | 0.07 | 0.06 | 0.08 | 0.07 | 0.08 | 0.08 | 0.09 | 0.10 | 0.11 | 0.12 | 0.14 | 0.16 |
| | | RF | **0.04** | **0.04** | **0.04** | **0.04** | **0.04** | **0.04** | **0.04** | **0.05** | **0.05** | **0.06** | **0.07** | **0.08** |

**Table 4**. Component Fragilities for Selected Bridge Class

| Components | Slight (LS$_1$) | | Extensive (LS$_2$) | | Extensive (LS$_3$) | | Complete (LS$_4$) | |
| --- | --- | --- | --- | --- | --- | --- | --- | --- |
| | $\lambda$ (g) | $\beta$ | $\lambda$ (g) | $\beta$ | $\lambda$ (g) | $\beta$ | $\lambda$ (g) | $\beta$ |
| Column | 0.175 | 0.350 | 0.479 | 0301 | 0.621 | 0.300 | 0.745 | 0.302 |
| Abutment passive | 0.824 | 0.296 | 1.640 | 0.251 | – | – | – | – |
| Abutment active | 0.651 | 0.299 | 1.114 | 0.245 | – | – | – | – |
| Abutment transverse | 0.199 | 0.303 | 0.520 | 0.316 | – | – | – | – |
| Bearing | 0.107 | 0.450 | 0.506 | 0.431 | – | – | – | – |
| Unseating | – | – | – | – | 4.419 | 0.612 | 20.275 | 0.881 |

**List of Figures**

**Fig. 1.** Illustration of a Regression Tree

**Fig. 2.** Numerical Modeling of Three-Span and Four-Span Seat-Type Abutment Bridge

**Fig. 3.** Development of Traditional Stripe-Based Fragility Curves

**Fig. 4.** (a) Histogram and (b) Probability Plot for Column Curvature Ductility at $S_{a-1.0}$ = 1.0g and (c) Histogram and (b) Probability Plot for Bearing Deformation at $S_{a-1.0}$ = 1.0g

**Fig. 5.** Comparison of Component Fragility Curves Using the Existing and Proposed Methods

**Fig. 6**. Relative Importance of Various Uncertain Parameters of Column Curvature Ductility Demand at $S_{a-1.0}$ = 0.2g, 0.6g, and 1.0g

**Fig. 7.** Relative Importance of Various Uncertain Parameters on All EDPs at $S_{a-1.0}$ = 0.6g

**Fig. 8.** Fragility Curves for Moderate Damage State

**Fig. 9.** Comparison of Actual and Fitted Fragility Curves for Selected Bridge Class

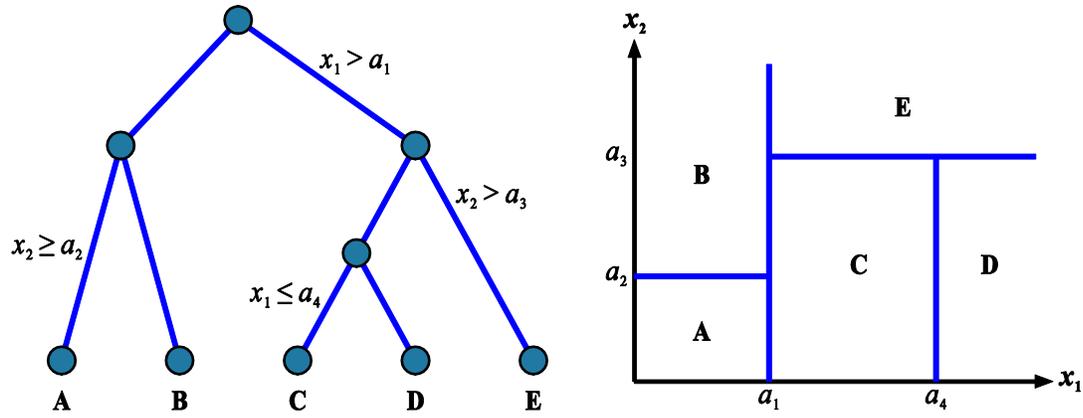

**Fig. 1.** Illustration of a Regression Tree

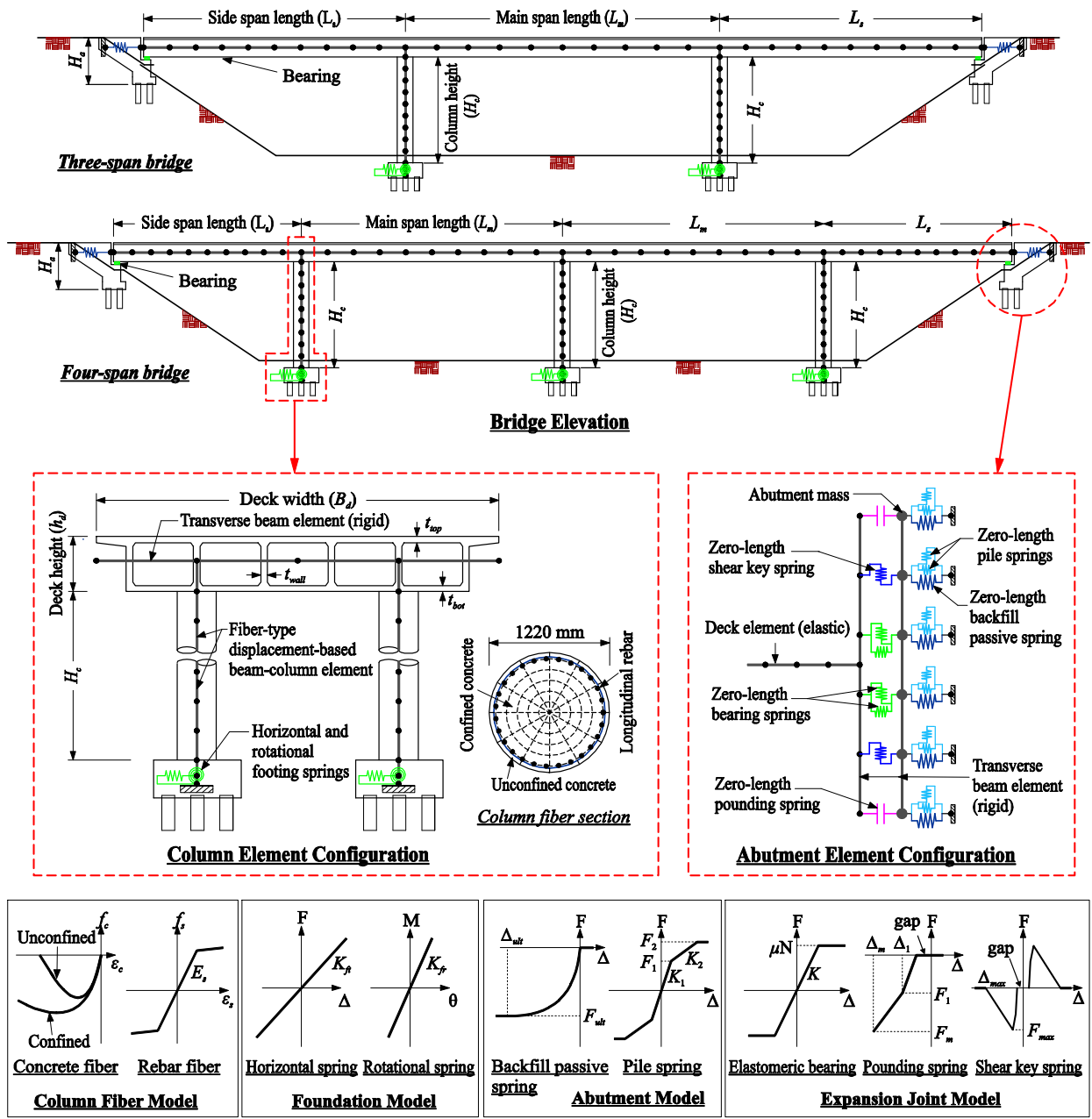

**Fig. 2.** Numerical Modeling of Three-Span and Four-Span Seat-Type Abutment Bridge

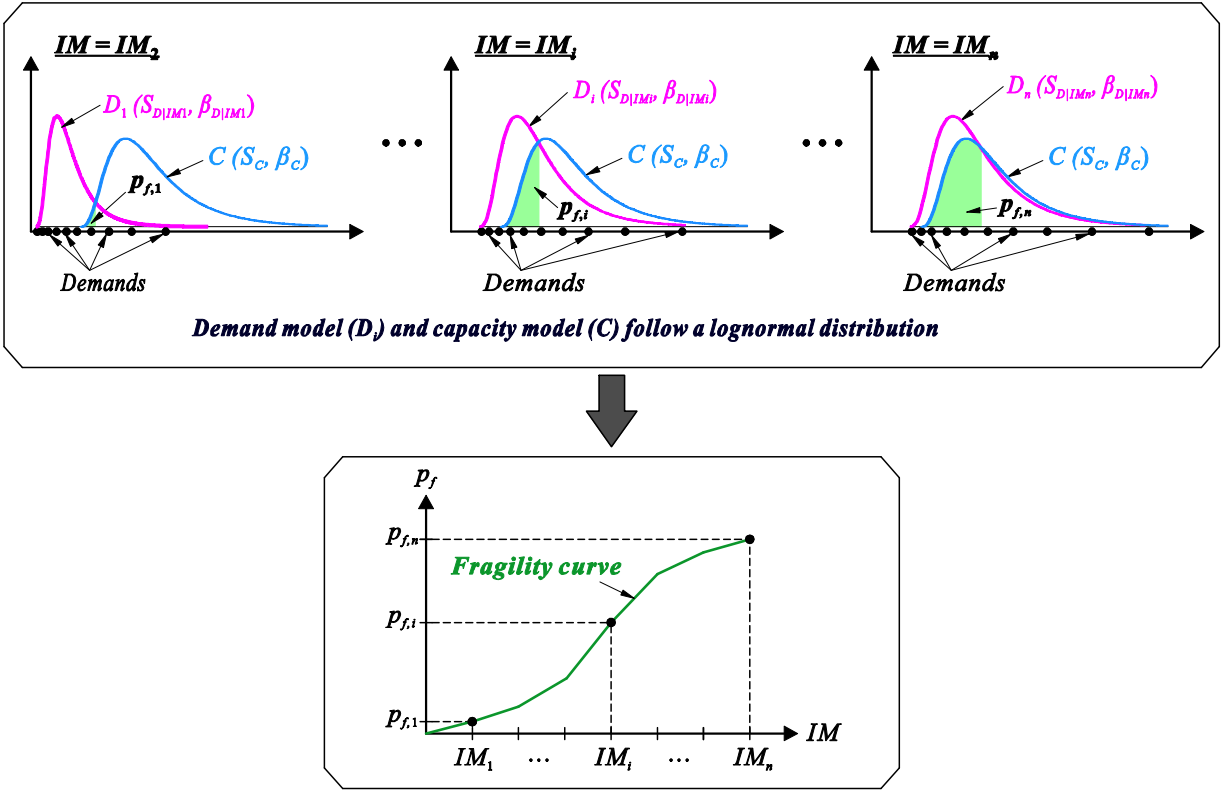

**Fig. 3.** Development of Traditional Stripe-Based Fragility Curves

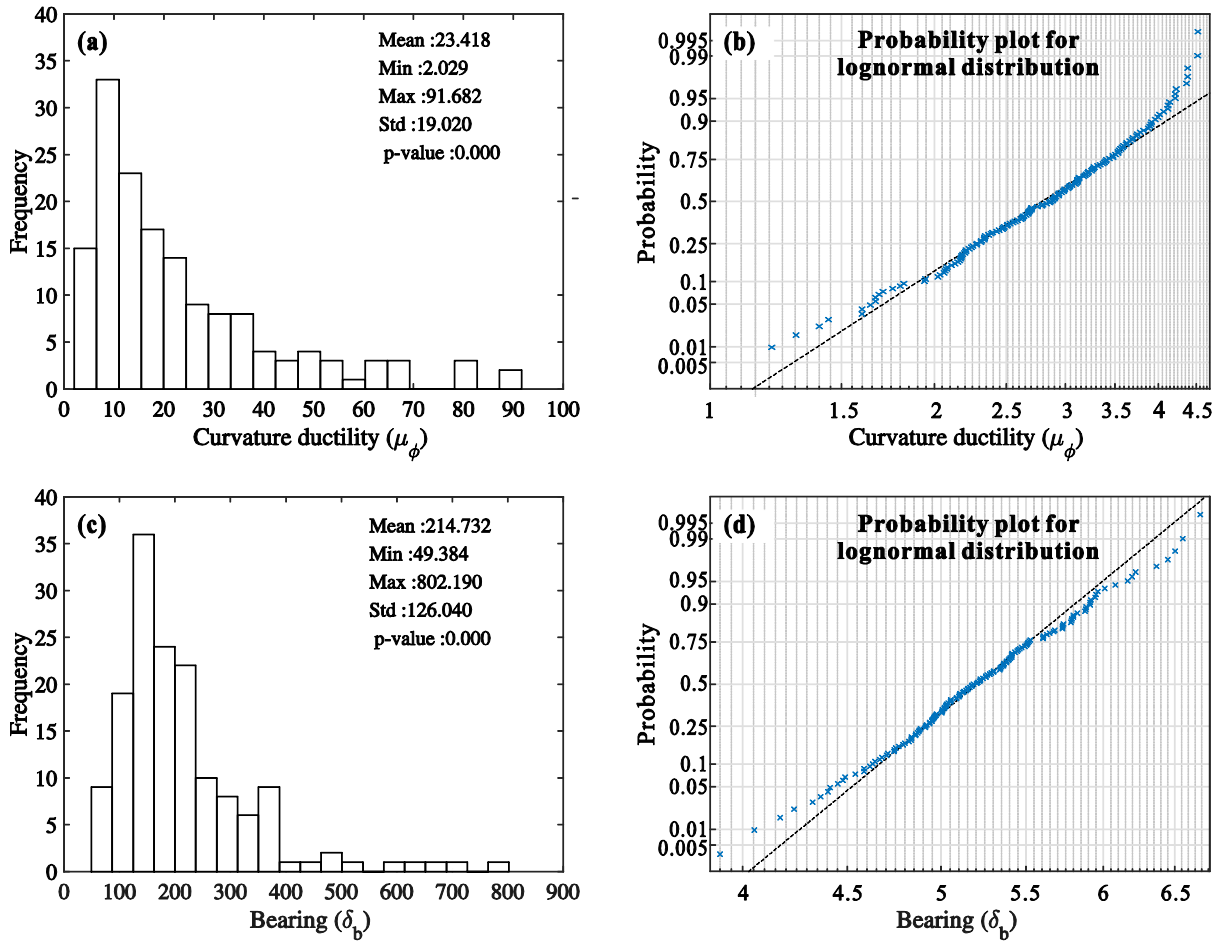

**Fig. 4.** (a) Histogram and (b) Probability Plot for Column Curvature Ductility at $S_{a-1.0}$ = 1.0g and (c) Histogram and (b) Probability Plot for Bearing Deformation at $S_{a-1.0}$ = 1.0g

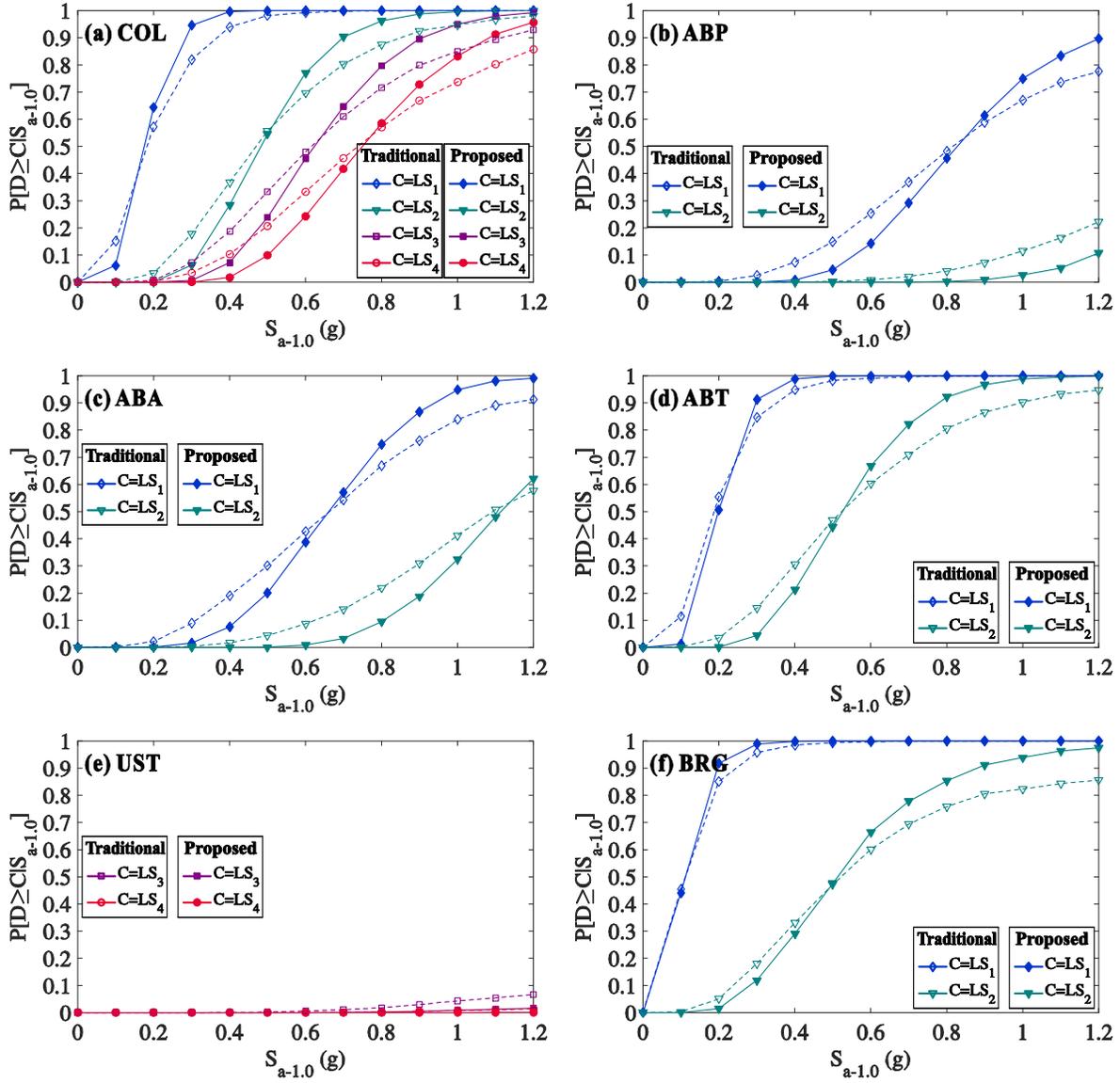

**Fig. 5.** Comparison of Component Fragility Curves Using the Existing and Proposed Methods

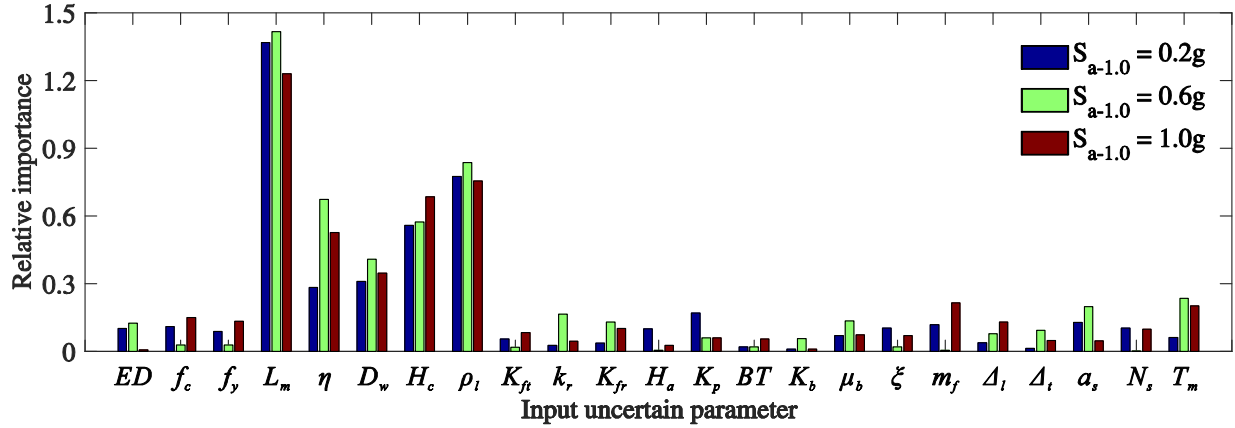

**Fig. 6.** Relative Importance of Various Uncertain Parameters of Column Curvature Ductility Demand at $S_{a-1.0}$ = 0.2g, 0.6g, and 1.0g

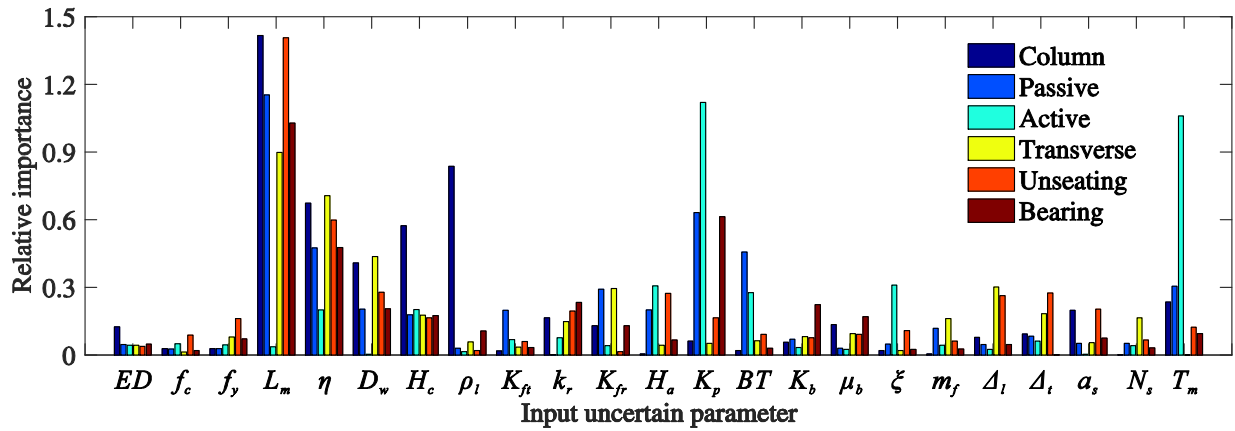

**Fig. 7.** Relative Importance of Various Uncertain Parameters on All EDPs at $S_{a-1.0}$ = 0.6g

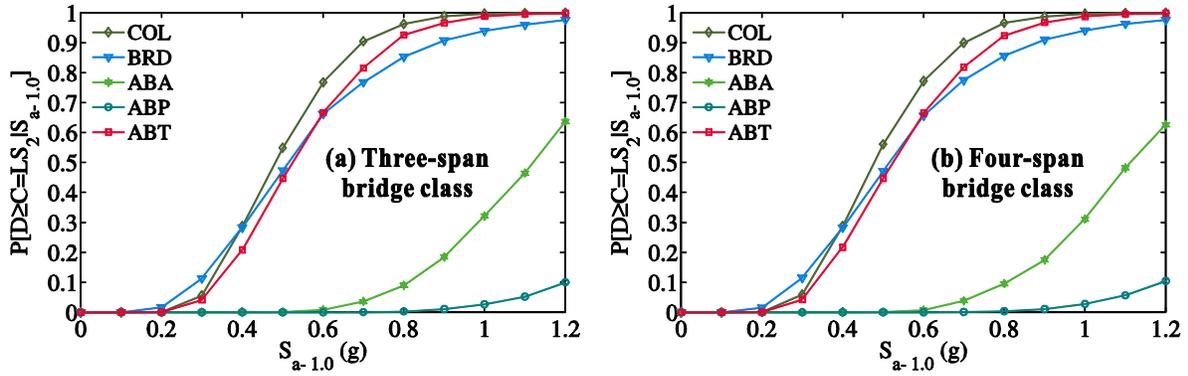

**Fig. 8.** Fragility Curves for Moderate Damage State

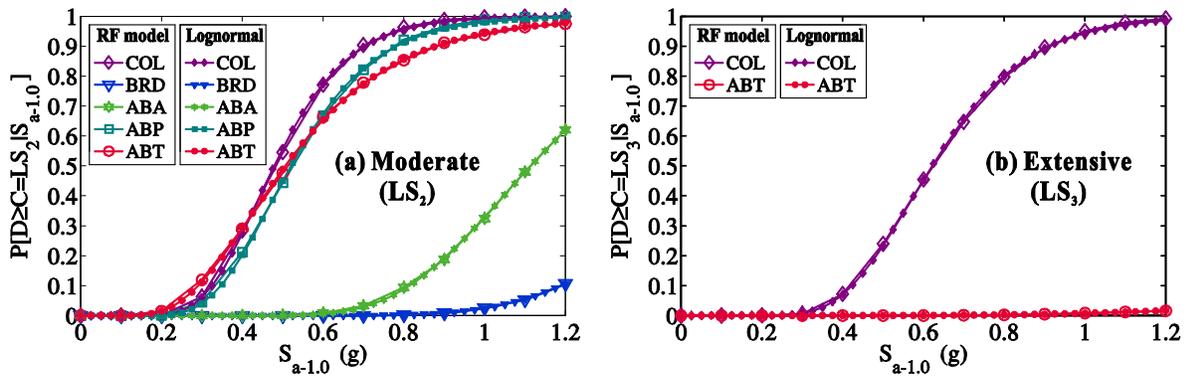

**Fig. 9.** Comparison of Actual and Fitted Fragility Curves for Selected Bridge Class